\documentclass[aps,subfigure,twocolumn]{revtex4}

\usepackage{graphicx}
\usepackage{dcolumn}
\usepackage{bm}
\usepackage{amsmath}
\usepackage{color,psfrag,epsfig}
\begin{document}

\newcommand{\bk}{{\bf k}}
\newcommand{\bc}{\begin{center}}
\newcommand{\ec}{\end{center}}
\newcommand{\mub}{{\mu_{\rm B}}}
\newcommand{\sD}{{\scriptscriptstyle D}}
\newcommand{\sF}{{\scriptscriptstyle F}}
\newcommand{\sCF}{{\scriptscriptstyle \mathrm{CF}}}
\newcommand{\sH}{{\scriptscriptstyle H}}
\newcommand{\sAL}{{\scriptscriptstyle \mathrm{AL}}}
\newcommand{\sMT}{{\scriptscriptstyle \mathrm{MT}}}
\newcommand{\sT}{{\scriptscriptstyle T}}
\newcommand{\up}{{\mid \uparrow \rangle}}
\newcommand{\down}{{\mid \downarrow \rangle}}
\newcommand{\upsp}{{\mid \uparrow_s \rangle}}
\newcommand{\downsp}{{\mid \downarrow_s \rangle}}
\newcommand{\upsone}{{\mid \uparrow_{s-1} \rangle}}
\newcommand{\downsone}{{\mid \downarrow_{s-1} \rangle}}
\newcommand{\upt}{{ \langle \uparrow \mid}}
\newcommand{\downt}{{\langle \downarrow \mid}}
\newcommand{\bbar}{{\mid \uparrow, 7/2 \rangle}}
\newcommand{\abar}{{\mid \downarrow, 7/2 \rangle}}
\renewcommand{\a}{{\mid \uparrow, -7/2 \rangle}}
\renewcommand{\b}{{\mid \downarrow, -7/2 \rangle}}
\newcommand{\plus}{{\mid + \rangle}}
\newcommand{\minus}{{\mid - \rangle}}
\newcommand{\psio}{{\mid \psi_o \rangle}}
\newcommand{\psis}{{\mid \psi \rangle}}
\newcommand{\bpsio}{{\langle \psi_o \mid}}
\newcommand{\barpsi}{{\mid \psi' \rangle}}
\newcommand{\barpsio}{{\mid \bar{\psi_o} \rangle}}
\newcommand{\ex}{{\mid \Gamma_2^l \rangle}}
\newcommand{\LH}{{{\rm LiHoF_4}}}
\newcommand{\LHx}{{{\rm LiHo_xY_{1-x}F_4}}}
\newcommand{\de}{{{\delta E}}}
\newcommand{\Ht}{{{H_t}}}
\newcommand{\sL}{{\cal{L}}}

\topmargin=-10mm

\title{$\LHx$ as a random field Ising ferromagnet}

\author{Moshe Schechter}
\affiliation{Department of Physics \& Astronomy, University of British Columbia,
Vancouver, B.C., Canada, V6T 1Z1}
\date{}

\begin{abstract}

As a result of the interplay between the intrinsic off-diagonal
terms of the dipolar interaction and an applied {\it transverse}
field $H_t$, the diluted $\LHx$ system at x$> 0.5$ is equivalent to
a ferromagnet in a longitudinal random field (RF). At low $H_t$ the
quantum fluctuations between the Ising like doublet states are
negligible, while the effective induced RF is appreciable. This
results in a practically exact equivalence to the classical RF Ising
model. By tuning $H_t$, the applied longitudinal field, and the
dilution, the Ising model can be realized in the presence of an
effective RF, transverse field, and constant longitudinal field, all
independently controlled. The experimental consequences for
$D=1,2,3$ dimensions are discussed.

\end{abstract}

\maketitle

Since its seminal discussion by Imry and Ma\cite{IM75}, the Ising
model in random longitudinal magnetic field was found to have many
interesting realizations in nature, and has been a subject of
intensive theoretical and experimental investigation for both
short-range and dipolar ferromagnetic (FM) interactions (see Refs.
\cite{Nat88,BY91,Bel98,Nat98} and references therein). In
particular, experiments are mostly done on diluted antiferromagnets
(DAFM)\cite{FA79}, which were shown to be equivalent in their static
critical behavior to the random field Ising model
(RFIM)\cite{FA79,Car84}. In the DAFM the order parameter is given by
the staggered magnetization, and a constant applied longitudinal
field plays the role of a tunable random field (RF). However, the
straight forward realization of the RFIM in a FM system has not
found an experimental realization, due to the difficulty of applying
a local random magnetic field.

Interestingly, the diluted $\LHx$ system provides the possibility of
producing effective local magnetic fields in the longitudinal
direction, due to the interplay of the off-diagonal terms of the
dipolar interactions and an applied constant magnetic field $\Ht$ in
the transverse direction\cite{Gin05,SS05,SL06,TGK+06}. Importantly,
as is further explained below, due to the strong hyperfine
interactions in this system, for $\Ht < 0.5$T the applied transverse
field does generate an effective longitudinal RF due to fluctuations
between each Ising ground state (GS) doublet and its corresponding
excited state\cite{SL06,SSL06}, but the coupling between the two
Ising doublet states is negligible\cite{SS05}.  Furthermore, despite
the presence of some correlations between the random fields (RFs),
and their origin in the long range dipolar interaction, we show
below that on length scales larger than the inter-vacancy distance
the behavior of the system is equivalent to that of the uncorrelated
RF model. Therefore, despite the presence of a finite $H_t$, and
finite correlations on short scales, $\LHx$ at $\Ht < 0.5$T {\it is
a perfect realization of the classical {\rm RFIM} with a tunable
random magnetic field.}

\begin{figure}
\includegraphics[width = 0.75\columnwidth]{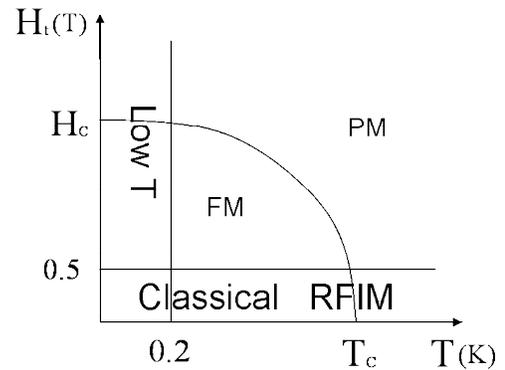}
 \caption{ Schematic picture of the phase diagram at some dilution x$<1$. At $\Ht < 0.5$T
 the quantum fluctuations between the Ising doublet states are negligible, while the effective random
 field is already appreciable. In this regime the system is equivalent to the classical RFIM.
 At $T \ll 0.2$K only the electro-nuclear ground states are
occupied, and the quantum
 phase transition in the presence of the random effective field can be studied.}
    \label{schematic}
\end{figure}

As $H_t$ is enhanced, the effective transverse field becomes
appreciable. Thus, the Ising model Hamiltonian
\begin{equation}
H=\sum_{ij} J_{ij} \tau_i^z \tau_j^z + \sum_i \gamma_i \tau_i^z +
\Delta \sum_i \tau_i^x + h_{\|} \sum_i \tau_i^z
 \label{fullIsing}
\end{equation}
can be realized in the $\LHx$ with an effective local RF term
$\gamma_i$, an effective transverse (quantum) term $\Delta$, and an
effective constant field in the longitudinal direction $h_{\|}$. In
the regime considered in this Rapid Communication, where x$>0.5$,
the system is FM at low $T,H_t$\cite{Ros96,BH07}, and $J_{ij}$
correspond to the long range dipolar interaction. In comparison to
the DAFM systems, the $\LHx$ system is different since in the former
the effective interaction is short range, and advantageous since in
the DAFM system one can realize neither an effective longitudinal
field in the staggered magnetization, nor an appreciable quantum
term. Importantly, the three effective fields, $\Delta, h_{\|}$ and
$\gamma \equiv \sqrt{{\it Var}(\gamma_i)}$ can be independently
controlled by tuning $H_t$, the applied magnetic field in the
longitudinal direction $H_z$, and the dilution x. Specifically, for
$H_t \ll 1$T and appreciable dilution $\gamma \gg \Delta$, where
$\Delta \gg \gamma$ for $H_t > 1.5$T and $\bar{\rm x} \equiv 1-{\rm
x} \ll 1$.

This gives for the first time an opportunity to study experimentally
the RFIM in a FM system, in both the classical and quantum regimes.
In particular, the realization of the Hamiltonian (\ref{fullIsing})
in a FM system allows both the study of long-standing questions
using direct bulk probes such as magnetization and susceptibility,
and for the first time an experimental study of models in which in
addition to the random field a constant longitudinal field or a
quantum term plays a role. Thus, the present work could be a basis
for numerous new experimental possibilities, a few of which are the
following: (i) The thermal and quantum phase transitions (PTs) of
the RFIM at $D=3$ can be studied in the whole $T, \Ht$ phase diagram
(see Fig.\ref{schematic}). Experiments studying the classical phase
transition in $3$D DAFM systems show that the RF prevents the sample
from reaching equilibrium on experimentally accessible time
scales\cite{BCSY85,YZM+06}. The study of the phase transition in
diluted dipolar magnets can shed light on this interesting
phenomenon, and the effect of the dipolar interaction on the
theoretical predictions for the equivalence of the quantum and
classical PTs in the RFIM\cite{AGS82,Sen98} can be tested as
well\cite{dipolarnote}. Of particular interest is the ability,
resulting from the novel quantum term, to control the relaxation
rate of the system by tuning the effective transverse field. (ii) As
a result of the possibility to apply tunable random field and
constant field in the longitudinal direction, one could study
experimentally the hysteresis in the RFIM as function of $H_z$.
Theoretical studies of this model in the absence\cite{SDM01} and
presence\cite{CD03} of long range interactions have shown an
intriguing disorder-driven out of equilibrium PT, a paradigm for the
study of crackling noise in self-organized critical systems. (iii)
In $D=1,2$ dimensions, the predictions of Imry-Ma\cite{IM75} and
Binder\cite{Bin83} for the destruction of long-range FM
order\cite{AW89} and the rounding of the phase transition in $2$D
can be verified by studying the magnetization as function of the
$\Ht$ dependent effective RF.

All our analysis and results below can be easily generalized to any
dipolar Ising magnet, as is explained in detail in Ref.\cite{SL06}.
In the following we concentrate on the $\LHx$ compound since this
enables us to present quantitative results for this system which is
of prime experimental interest.

{\it Realization of the {\rm RFIM} in the $\LHx$.---} Anisotropic
dipolar magnets in general, and the $\LHx$ compound in particular,
are considered to be realizations of the Ising model. As a result of
the crystal field anisotropy the GS is an Ising like doublet, and
all but the longitudinal dipolar interactions are effectively
reduced. The application of $\Ht$ induces quantum fluctuations (QF)
between the Ising ground states (GSs), resulting in an effective
transverse field Ising model (TFIM). For the undiluted system both
the thermal and the quantum PT to the PM state were
observed\cite{BRA96}. The dominant interaction in the $\LHx$ system
is dipolar\cite{CHK+04} $H_{\rm dip} = \sum_{ij,\alpha \beta}
V_{ij}^{\alpha \beta} J_i^{\alpha} J_j^{\beta}$. In the pure $\LH$,
the off-diagonal terms of the dipolar interaction, e.g. $J_i^z
J_j^x$, are not only effectively reduced, but cancel by symmetry.
However, at finite dilution, x$<1$, this cancelation does not hold.
Furthermore, in Ref.~\cite{SL06} it was shown that in the presence
of a constant $\Ht$, the off-diagonal terms, despite being
effectively reduced, are significant since their presence changes
the symmetry of the system. The system is no longer symmetric under
the transformation $S_z \rightarrow -S_z$, and an effective local
longitudinal magnetic RF $\tilde{h}_j$ is generated. In the regime
$H_t \ll \Omega_0/(\mub S)$

\begin{equation}
\tilde{h}_j = \frac{2S \Ht}{\Omega_0} \sum_iV_{ji}^{zx} \label{hj} ;
\end{equation}
see Ref.\cite{SL06}. The random local energy is given by
$\tilde{\gamma}_j=\mub \tilde{h}_j S$. $\Omega_0 \approx 10$K is the
measure of the anisotropy, given by the excitation energy between
each of the Ising GSs to its relevant excited state\cite{SL06}, and
$S \approx 5.5$ is the magnitude of the effective Ho spin.

\begin{figure}
\psfrag{N}{$N$} \psfrag{IM}{$\Sigma^2/\bar{\rm x}^2$}
\psfrag{NUM}{$\bar{\rm x}$} \psfrag{IMRY}{$\Sigma/\sqrt{N}$}
\includegraphics[width = \columnwidth]{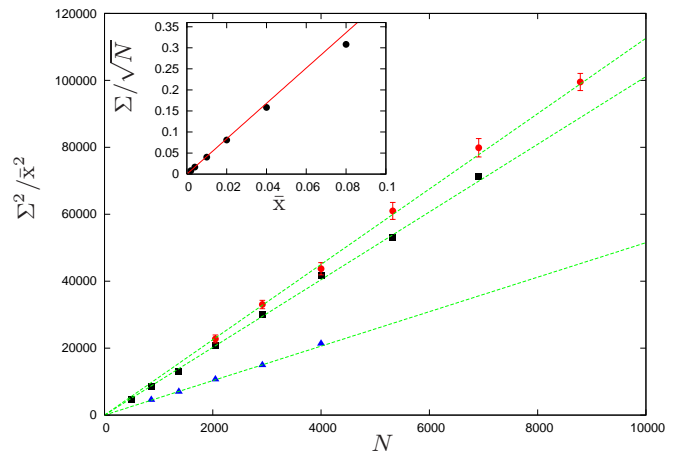}
\caption{$\Sigma^2/\bar{\rm x}^2 \equiv (E_{IM}/c V_0 \bar{\rm
x})^2$ in three dimensions is plotted as function of $N=4*L^3$ for
dilutions x$=0.6$ (blue triangles), $0.96$ (black squares), and
$0.996$ (red circles). $L$ is the length of the cube in unit cells.
Averaging is done over $10^4$ samples for each data point. Where not
drawn, errorbars are smaller than the point size. Inset:
$|\Sigma|/\sqrt{N}$ as function of $\bar{\rm x}$ for $N=4000$.
$|\Sigma| =\eta_o \bar{\rm x} \sqrt{N}$ for $\bar{\rm x} \ll 1$,
with $\eta_o=4.12$.} \label{figure3d}
\end{figure}

Importantly, $\tilde{h}_j$ (\ref{hj}) is independent of the
directions of the spins at sites $i$, and therefore equivalent for
the spin-glass (SG) and FM regimes. However, the energy gain due to
the effective RF does depend on the spin configuration. Thus, while
the randomness in the energy gain in the SG regime is a result of
both the random position and the random orientation of the spins, in
the FM regime only the former are random. Furthermore, the RFs
$\tilde{h}_j$ are not uncorrelated. For $\bar{\rm x} \ll 1$ the
correlations of $\tilde{h}_j$ at distances smaller than the typical
inter-vacancy distance, are large. However, as is shown explicitly
below, for all dilutions, for distances larger than the
inter-vacancy distance, the typical energy gain in flipping a domain
$E_{IM} = \sum_j \tilde{\gamma}_j$ obeys the Imry-Ma behavior. In
particular ${\it Var}(E_{IM}) \propto N$, where $N$ is the number of
sites in a domain.

Let us now calculate explicitly $\Sigma^2 \equiv (\sum_{ij}
V_{ij}^{zx}/V_0)^2 = (1/c V_0)^2 E_{IM}^2$, where $c \equiv 2 \mub
H_t S^2/\Omega_0$ and $V_0$ is the magnitude of the dipolar
interaction at distance of one unit cell along the $x$ direction.
First, note that for a single vacancy $\bar{i}$, as a consequence of
lattice symmetry, $\sum_j V_{\bar{i} j}^{zx}=0$. This results in
$\sum_{ij} V_{ij}^{zx} = \sum_{\bar{i} \bar{j}} V_{\bar{i}
\bar{j}}^{zx}$, where $i,j$ run over occupied sites and $\bar{i},
\bar{j}$ over the unoccupied sites. Consider first the case of
$\bar{\rm x} \ll 1$. Assuming that the interactions causing the RF,
$V_{ij}^{zx}$, are nearest neighbor only, then in leading order
$E_{IM}^2$ is proportional to the number of nearest neighbor vacancy
pairs, and therefore to $\bar{\rm x}^2$. In analogy to the case of
dipolar interaction in a random system\cite{BMY86} it is expected
that {\it for the random interaction leading to the effective RF}
the dipolar and short range interactions would be equivalent. Thus,
we expect ${\it Var}(E_{IM}) \propto N$ for all dilutions and ${\it
Var} (E_{IM}) \propto \bar{\rm x}^2 N$ for $\bar{\rm x} \ll 1$,
where correlations of three or more vacant spins can be neglected.

In Fig.\ref{figure3d} we plot ${\it Var}(\Sigma)/\bar{\rm x}^2$ for
cubes in $3$D as function of $N=4L^3$ for x$=0.6,0.96,0.996$. Here
and below lengths are given in terms of lattice unit cells. The
dilutions were chosen to include the cases where $\bar{\rm x} \ll
1$, and effective fields are strongly correlated on small distances,
x$=0.6$ where correlations due to the relative positions of three or
more vacancies are significant, and x$=0.96$ in between. For each x
we obtain a good linear fit, ${\it Var}(\Sigma)/\bar{\rm x}^2 =
(\eta({\rm x}))^2 N$, which corresponds to ${\it Var}(E_{IM}) =
(\eta({\rm x}) c \bar{\rm x} V_0)^2 N$. In the inset we plot
$\sqrt{{\it Var}(\Sigma)/N}$ as function of $\bar{\rm x}$ for a $3$D
cube with $N=4000$, showing $\lim_{\bar{\rm x} \rightarrow 0} \eta =
\eta_o$ with $\eta_o = 4.12$.

We therefore conclude that for distances larger than the
inter-vacancy distance the system is equivalent to a RFIM, with an
effective {\it uncorrelated} RF

\begin{equation}
\gamma_j = \eta_j \frac{2 \mub H_t S^2}{\Omega_0}V_0 \bar{\rm x} ,
\label{gammaj}
\end{equation}
$\eta_j$ being a random number with $\langle \eta_j \rangle = 0$ and
${\it Var}(\eta_j) = \eta^2/{\rm x}$. Similar results (not shown)
were obtained for $D=1$. Interestingly, $\gamma_j$ in
Eq.(\ref{gammaj}) is similar to the RF obtained in the SG
regime\cite{SL06}, with $\bar{\rm x}$ replacing x and the constant
prefactor being different for the two regimes due to the different
spin configuration.

While our results are general to any dipolar Ising magnet, we now
use the specifics of the $\LHx$ system to show that despite the
presence of $H_t$ the analogy to the classical RFIM is possible. The
Ising like anisotropy is dictated by a crystal field Hamiltonian
having large $J_z^2$ and $J_z^4$ terms\cite{CHK+04}. In addition, a
large $(J_+^4 + J_-^4)$ term results in a mixing of free ion states
with $\Delta J_z = \pm 4$. The electronic GSs are an Ising like
doublet, denoted $\up, \down$, belonging to the $J_z=(\pm7, \pm3,
\mp1, \mp5)$ multiplets. Although the full splitting of the
$2J+1=17$ states is $\Omega \approx 600$K, the first excited state
$\mid \Gamma \rangle$, belonging to the $J_z=(6,2,-2,-6)$ multiplet
is only $\Omega_0=10$K higher than the GS doublet. The electronic
Hamiltonian allows a coupling between the electronic GSs which is
second order in $\Ht$. However, the low energy physics of the $\LHx$
system is strongly affected by the strong hyperfine (hf) interaction
between the Ho electronic angular momentum and its nuclear
spin~\cite{SS05,SS07}. In particular, each electronic GS is split to
$2I+1=8$ nearly equidistant states with $I_z = -7/2, \dots , 7/2$
and a separation of $205$mK between them\cite{GWT+01}. Thus, the
relevant Ising like single spin GS doublet is of electronuclear
type, having a definite and opposite nuclear spin states, i.e. $a
\equiv \a$ and $\bar{a} \equiv \abar$. The tunneling between the
states $a$ and $\bar{a}$ requires the flipping of the nuclear spin,
resulting in a very weak coupling at low fields [less than
$10^{-4}$K at $\Ht < 1$T (Ref. \cite{SS07})]. Only when $\mid \Gamma
\rangle$ becomes well hybridized with the electronic GSs $\up,
\down$, QF between the states $a$ and $\bar{a}$ become appreciable.
This occurs at $\Ht \approx \Omega_0/(\mub S) \approx 1.5$T (see a
detailed plot of $\Delta(H_t)$ in Fig 2 of Ref.\cite{SS05}).

We first consider the low $T$ low $\Ht$ (LTLH) regime ($\Ht < 1$T
and $T \ll 0.2$K). In this regime only the states $a$, $\bar{a}$ are
appreciably occupied, and the quantum tunneling between these two
states is negligible. Thus, the system is effectively a classical
Ising model\cite{SS05}. Importantly, the effective
RF~(\ref{gammaj}), originating from fluctuations between each Ising
GS and its corresponding excited state\cite{SL06,SSL06}, can already
be made appreciable, since its behavior is linear in $\Ht$. Thus, in
this regime, which is of particular interest for the study of the
destruction of long range order (LRO) by the effective RF in $D=1,2$
(see below), and the equilibrium and nonequilibrium phenomena in the
ordered state in $D=3$, {\it the analogy between the $\LHx$ system
and the {\rm RFIM} is practically exact}.

{\it Phase transitions in three dimensions.---} As a result of the
very different dependence of $\gamma$ and $\Delta$ on $H_t$, the
thermal and quantum PTs can be well separated, as is sketched in
Fig.\ref{schematic}. Both phase transitions occur outside the LTLH
regime. For the classical PT one has to go to $T>0.2$K, since $T_c
\approx {\rm x}*1.54$K\cite{Ros96}. Therefore, all nuclear states
are appreciably occupied near the transition. Still, at low enough
transverse field, $\Ht \lesssim 0.5$T, QF are small between any of
the time reversed states $\mid \uparrow, -I_z \rangle, \mid
\downarrow, I_z \rangle$~\cite{SS07}. Thus, the nuclear spin states
can be traced over, and the classical PT is equivalent to that of
the RFIM. Note, that the magnitude of the RF can be tuned by varying
the dilution, and for each given sample by changing the magnitude of
$H_t$, as can be inferred from Eq.~(\ref{gammaj}).

Considering the quantum phase transition (QPT) we focus on $T \ll
0.2$K, where only the GS doublet is appreciably occupied. However,
we still have to go out of the LTLH regime since strong QF are
required. Specifically, in the dilutions of interest the QPT occurs
at $H_t > 3$T. In Ref.\cite{SS05} it was shown that QF between the
electronuclear Ising states become appreciable at $\Ht \approx
1.5$T, and at $\Ht \approx 3$T the GS has all hf levels well
mixed~\cite{SS07}. However, since at x$>0.5$ the typical magnitude
of the dipolar energy $V_0 {\rm x} S^2 \gg 0.2$K, it is only when
the QF of the {\it electronic} spins are larger than the interaction
that the PT to the PM phase takes place. At x$=1$ it was argued that
the system is equivalent to the TFIM, with the hf interactions
effectively re-normalizing the transition field~\cite{CHK+04}. For
x$<1$ and $V_0 {\rm x} S^2 \gg 0.2$K we expect the same picture to
be valid, with the addition of the effective longitudinal RF.

At the QPT $\Ht \approx \Omega_0/(\mub S)$, Eq.~(\ref{hj}) is not
valid, and QF between the Ising doublet states enhance the effective
RF. The largest RFs that can be obtained are of the order of $V_0
S/\mub$. By changing the dilution x, one can therefore control the
magnitude of the typical random magnetic field $h$ at the QPT in the
regime $0 < h \lesssim V_0 S/\mub$, equivalent to $0 < \gamma
\lesssim J$, where $J$ is the average nearest neighbor interaction.
Thus, for $\bar{\rm x} \ll 1$ the QPT can be studied in the presence
of weak RF, while for a substantial dilution the RF, as well as the
fluctuations, are of the order of the interaction. Note, that LRO is
destroyed when either QF or the RF are of the order of the
interaction, and the situation where all three energy scales are of
the same order is of both theoretical and experimental interest.

{\it Measuring the correlation length in one and two dimensions.---}
Two is the critical dimension for the stability of the FM phase to
the RF\cite{IM75,Nat88}, as the energy gain and energy cost to flip
a domain are both linear in the domain size. However,
Binder\cite{Bin83} has shown that due to the roughening of the
domain wall, a two dimensional FM Ising system in a RF has a finite
correlation length $\xi^{2D} \propto \exp{(V_0 S^2/\gamma)^2}$. Here
we consider first a strictly 2D system, where the plane includes the
easy $z$-axis and is chosen as to have a FM phase at zero transverse
field. We first consider an in-plane magnetic field $H_x$ transverse
to the easy axis. For $H_x \ll \Omega_0/(\mub S)$ the effective RF
is given by Eq.~(\ref{gammaj}), and therefore, for an appreciable
dilution, say $0.5<{\rm x}<0.8$, $\xi^{2D} \propto
\exp{[(\Omega_0/\mub H_x)^2]}.$ Since $S \gg 1$, $\xi^{2D}
> \exp{(S^2)}$ in the perturbative regime. As $H_x$ is increased,
QF between the Ising like doublet become significant, $h$ increases
and $\xi$ decreases. Assuming one can reach the regime where
$\xi^{2D} \ll L$, the finite correlation length can be detected by
studying $M$ as function of $H_x$

\begin{equation}
M \approx M_0(H_x) \frac{\xi^2}{L^2} \sqrt{\frac{L^2}{\xi^2}}
\approx M_0(H_x) {\frac{\xi}{L}} \, , \label{mag2d}
\end{equation}
as the ratio $M/M_0$ depends linearly on $\xi$. Here $M_0(H_x)$
corresponds to the dependence of the magnetization on $H_x$ in a FM
domain due to single spin fluctuations, as can be measured in a $3$D
sample where the system stays FM for the relevant magnetic fields
[$\Ht \ll \Omega_0/(\mub S)$].

Unlike systems such as the DAFM ${\rm
RB_2Co_xMg_{1-x}F_4}$\cite{FKJ+83}, where the two dimensionality is
due to the inherent layered structure, for the $\LHx$ a realistic
system would be quasi two-dimensional, with a finite width $d$. The
energy cost to flip a domain is then linear in $d$, while the energy
gain is $\propto \sqrt{d}$. As a result $\xi$ is enhanced,
$\xi^{Q2D} \propto \exp{[\Omega_0^2 d/(\mub H_x)^2]}$, and therefore
one has to keep $d$ small in order to observe a finite $\xi$. In
comparison to layered systems, the observation of a finite $\xi$ is
more difficult. However, the study of the gradual cross-over between
two and three dimensions is made possible. Note, that the degree of
two dimensionality can be studied by measuring the dependence of the
RF on the angle of $H_t$ in the $xy$ plane. This also allows the
continuous change of the RF in a given sample. In particular, for
the strict 2D case there is no RF when $H_t$ is perpendicular to the
plane.

Using simple Imry-Ma\cite{IM75} arguments one can show that for a
quasi-$1$D system parallel to the $z$-axis and of cross-section $A$,
the finite correlation length is given by $\xi^{1D} \approx
A(\Omega_0/\mub H_t)^2$, and the magnetization by

\begin{equation}
M \approx M_0(\Ht) \frac{\xi}{L} \sqrt{\frac{L}{\xi}} \approx
M_0(\Ht) \sqrt{\frac{A}{L}} \frac{\Omega_0}{\mub \Ht} \, .
\label{mag1d}
\end{equation}
Thus, the observation of $M/M_0 \propto 1/\Ht$ in the regime
$\Omega_0 \sqrt{A/L} \ll \mub \Ht \ll \Omega_0/S$ is the
experimental manifestation of the instability of the FM phase in one
dimensional Ising system to RF, as predicted by Imry and Ma.

As explained in detail in Refs. \cite{SL06,SSL06}, our analysis
above applies directly to a general anisotropic dipolar system, with
a crystal field Hamiltonian e.g of the form $H_{CF}=D S_z^2$, and
with no hyperfine interactions. Interestingly, in the $\LHx$ system,
if one neglects the hyperfine interactions, the peculiar form of the
crystal field Hamiltonian which allows transitions between the two
electronic ground states in second order perturbation, leads to
different physical results. Most importantly, the effective
transverse term becomes appreciable at small $\Ht$, and the RFIM is
obtained at the expense of significant quantum fluctuations. Thus,
the proper consideration of the hyperfine interactions in the $\LHx$
system is not only crucial for obtaining the correct effective
fields, but also to obtain the qualitative equivalence of the $\LHx$
system to general diluted anisotropic dipolar magnets.

It is a pleasure to thank Gabriel Aeppli, Amnon Aharony, David
Belanger, Nicolas Laflorencie, and Alessandro Silva for useful
discussions. This work was supported by NSERC of Canada and PITP.


\begin{thebibliography}{10}
\bibitem{dipolarnote}
The dipolar nature of the interaction has no effect on the validity
of the Imry-Ma arguments and on the lower critical dimension, but
its presence affects the critical behavior and breaks the
universality of the DAFM and RFIM systems, both in their low T
phases and the characteristics of the critical behavior\cite{Nat88}.
\end{thebibliography}
\end{document}